# Important factors for cell-membrane permeabilization by gold nanoparticles activated by nanosecond-laser irradiation

Cuiping Yao[1,2,*]
Florian Rudnitzki[2,*]
Gereon Hüttmann[2,3]
Zhenxi Zhang[1]
Ramtin Rahmanzadeh[2]

[1]Key Laboratory of Biomedical Information Engineering of Education Ministry, Institute of Biomedical Analytical Technology and Instrumentation, School of Life Science and Technology, Xi'an Jiaotong University, Xi'an, China; [2]Institute of Biomedical Optics, University of Lübeck, Lübeck, [3]Airway Research Center North (ARCN), Member of the German Center for Lung Research (DZL), Kiel, Germany

*These authors contributed equally to this work

**Purpose:** Pulsed-laser irradiation of light-absorbing gold nanoparticles (AuNPs) attached to cells transiently increases cell membrane permeability for targeted molecule delivery. Here, we targeted EGFR on the ovarian carcinoma cell line OVCAR-3 with AuNPs. In order to optimize membrane permeability and to demonstrate molecule delivery into adherent OVCAR-3 cells, we systematically investigated different experimental conditions.
**Materials and methods:** AuNPs (30 nm) were functionalized by conjugation of the antibody cetuximab against EGFR. Selective binding of the particles was demonstrated by silver staining, multiphoton imaging, and fluorescence-lifetime imaging. After laser irradiation, membrane permeability of OVCAR-3 cells was studied under different conditions of AuNP concentration, cell-incubation medium, and cell–AuNP incubation time. Membrane permeability and cell viability were evaluated by flow cytometry, measuring propidium iodide and fluorescein isothiocyanate–dextran uptake.
**Results:** Adherently growing OVCAR-3 cells can be effectively targeted with EGFR-AuNP. Laser irradiation led to successful permeabilization, and 150 kDa dextran was successfully delivered into cells with about 70% efficiency.
**Conclusion:** Antibody-targeted and laser-irradiated AuNPs can be used to deliver molecules into adherent cells. Efficacy depends not only on laser parameters but also on AuNP:cell ratio, cell-incubation medium, and cell–AuNP incubation time.

## Introduction

Targeted delivery and controlled release of therapeutic drugs to a specific cellular site is of great interest for basic research and clinical approaches. However, the efficiency of molecule delivery into cells still requires improvement.[1] Light-activated techniques allow for high spatial and temporal control of effects. The interaction of the light-absorbing gold nanoparticles (AuNPs) with short laser pulses leads to a localized increase in cell permeability for enhanced molecule delivery. This increase in permeability is transient, and the cell membrane reseals within 1 hour after irradiation.[2] Colloidal AuNPs have been investigated in biomedical research for cell inactivation, tumor treatment,[3,4] and nanosensing by tracking of cancer cells.[5,6] Further studies include targeted photothermal and photodynamic therapies,[7,8] AuNP-mediated radiation therapy,[9] in vitro biological analysis,[10] and molecule delivery into cells.[11] Extensive research has been implemented for cancer-cell killing by targeted drug delivery.[12–16] AuNPs have their absorption peak at around 520 nm, which enables efficient heating

Correspondence: Ramtin Rahmanzadeh
Institute of Biomedical Optics, University of Lübeck, 4 Peter-Monnik-Weg Street, Lübeck 23562, Germany
Tel +49 451 3101 3210
Fax +49 451 3101 3204
Email rahmanzadeh@bmo.uni-luebeck.de

of the particles by pulsed-laser irradiation to more than 1,000 K. To achieve thermal confinement to a radius of less than 100 nm in water, the pulse duration should be shorter than 10 nanoseconds.[17]

Different light sources and different AuNP sizes have been used to implement molecule delivery into cells. Differences in induced membrane-permeabilization behavior between nanosecond and picosecond lasers have been observed.[18] Cell permeabilization with AuNPs, where irradiation was shifted to longer wavelengths from their absorption peak at 800 nm, also referred to as off-resonant irradiation, has been demonstrated with a femtosecond laser.[19] Based on this method, the fluorescent dye Lucifer yellow YFP-Smad2 cDNA plasmid was delivered into cells with a high perforation rate of 70% and low toxicity (1%). Also, differences in membrane permeabilization by on- (532 nm) and off-resonance (1,064 nm) laser illuminations were compared.[20] The results showed that both lasers with different wavelengths were able to induce membrane permeabilization, but irradiation with near-infrared pulses offer better reproducibility and higher optoporation efficiency than those obtained with 532 nm pulses. With carbon NPs activated by a femtosecond laser, the delivery of calcein molecules into corneal endothelial cells with median efficiency as high as 54.5% and mortality as low as 0.5% has been shown.[21] Another effective transfection technique is based on laser scanning of cells that have been incubated with AuNPs, named GNOME (gold nanoparticle-mediated) laser transfection, and demonstrated the delivery of green fluorescent proteins into mammalian cells with an efficiency of 43% and high cell viability.[1] This technique combines high-throughput transfection of about 10,000 cells/second with a high cell-survival rate. In addition to the aforementioned techniques, other approaches, such as plasmonic nanobubble generation under laser irradiation[22] and laser-induced shockwave generation, have also been used to deliver molecules[23] or transfect cells in vivo and in vitro.[24]

In earlier studies, we showed the delivery of macromolecules like fluorescein isothiocyanate (FITC)–dextran or antibodies into the suspension cell lines Karpas299 and L428 using 30 nm AuNPs activated by nanosecond-pulsed laser.[2] Although different irradiation parameters, including nanosecond,[2,20] picosecond,[1,18] and femtosecond pulses,[19,21] and different AuNP sizes (30, 100, and 200 nm) with different concentrations have been used for achieving targeted transfection, an optimization study for adjusting those parameters is important for maximizing transfection efficiency. Adherent cells were used as target cells in all these studies, except Lukianova-Hleb et al[22] and our study.[2] However, in the former, single laser pulses were focused on individual cells, while a large number of cells were irradiated with scanning in our study.

To target the adherently growing cell line OVCAR-3, we used Au conjugated with the antibody cetuximab, directed against EGFR. The transmembrane protein EGFR is overexpressed on the ovarian carcinoma cell line OVCAR-3. Cetuximab conjugation leads to close localization of AuNPs at the cell membrane. Moreover, it adds selectivity for EGFR-overexpressing cell lines. The localization and selective binding of the conjugates were investigated with silver enhancement, immunofluorescence imaging, and fluorescence-lifetime imaging. After nanosecond-laser irradiation, we found permeabilization of the adherent cells and further that transfection efficacy depends not only on laser parameters but also on AuNP:cell ratio, cell-incubation medium (culture medium or PBS), and cell–AuNP incubation time.

## Materials and methods
### Laser system and irradiation platform

The laser-irradiation setup is shown in Figure 1. A frequency-doubled, Q-switched neodymium-doped yttrium aluminum garnet laser (Surelite I-20; Continuum, Santa Clara, CA, USA), which generates 4-nanosecond pulses at 532 nm, was used for irradiating samples to induce surface-plasmon resonance of AuNPs. The laser-pulse energy can be adjusted through a broadband half-wave ($\lambda/2$) plate and a polarizing beam splitter. The laser beam is directed perpendicularly onto the sample using a reflecting prism. The irradiation spot size on the sample surface can be adjusted with a focusing lens. The cell samples are placed in a 96-well plate held by a customized plate holder, which is fixed on an $x$-$y$-motorized stage. The samples are moved in a meander pattern through the laser beam.

### Cell culture

Human ovarian cancer cell line OVCAR-3 were obtained from the American Type Culture Collection. Cells were cultured in RPMI 1640 complete medium that was modified to contain 10 mM HEPES, 1 mM sodium pyruvate, 4.5 g/L glucose, 1.5 g/L sodium bicarbonate, and 0.01 mg/mL bovine insulin. The cells are cultured in a humidified atmosphere of 5% $CO_2$ at 37°C.

### AuNP–antibody conjugation and stability evaluation
#### Gold-nanoparticle functionalization

The anti-EGFR antibody cetuximab (10 µg/mL) was conjugated to 30 nm AuNPs through noncovalent electrostatic bonding. AuNPs with optical density OD of 1 at 523 nm in suspension were obtained from British Biocell International (Cardiff, UK). For effective conjugation of the antibody cetuximab, the pH of the suspension was adjusted to 7.5 with titration of $K_2CO_3$. Saturation of antibody linkage was determined through the incubation of different antibody concentrations with AuNPs, followed by a stability test using a 1.7 M sodium chloride solution. Ultraviolet-visible absorption spectra were measured with spectrophotometry (SpectraMax M4 [Molecular Devices, Sunnyvale, CA, USA] or U-2900 [Hitachi, Tokyo, Japan]) to determine the saturation concentration of the antibody. A color shift to blue indicated aggregation of AuNPs due to insufficient antibody binding. For experiments, the lowest antibody:AuNP ratio without broadness of absorption peak after addition of sodium chloride was used.

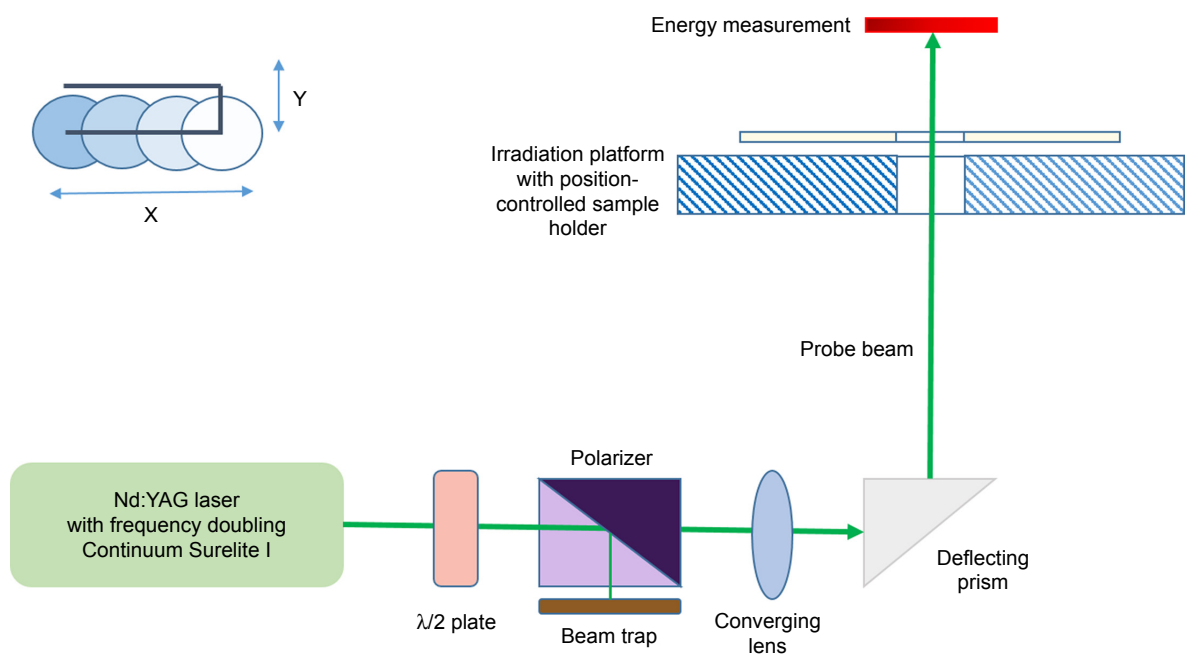

**Figure 1** Laser-irradiation system. The scanning pattern of the irradiation beam is shown in the upper-left corner.
**Abbreviation:** Nd:YAG, neodymium-doped yttrium aluminum garnet.

### Incubation of cells with AuNPs and testing for specific binding

OVCAR-3 cells were seeded into a 96-well plate at $2.5 \times 10^4$ cells per well with 200 µL medium. After 24-hour incubation, the medium was removed and the cells washed with PBS containing calcium and magnesium. Then, the cells were incubated with the AuNP–antibody conjugates (50 µL) for 20 minutes at 37°C. After incubation, the samples were washed again with PBS.

### Silver enhancement

Besides spectral analysis, successful biofunctionalization and localization of AuNPs at cell membranes were verified by microscopic observation using a silver-enhancement assay (British Biocell International). In this test, two reagents – initiator and enhancer (one drop each) – were added to OVCAR-3 cells incubated with AuNP conjugates. In the presence of gold, the silver nitrate of the solution is reduced to free silver, which settles on the gold surface and makes AuNPs visible for light microscopy. Two control samples of cells, one without AuNPs and the other incubated with nonfunctionalized AuNPs, were prepared. At 30 minutes after the silver-enhancement reagents were added, cell samples were observed under microscopy (Eclipse TE2000; Nikon, Tokyo, Japan).

### Detection with fluorescence-labeled antibody

For detection of antibodies with a secondary antibody, OVCAR-3 cells were incubated with the goat antihuman IgG Alexa Fluor 546 antibody, which was added to the cell petri dish at a concentration of 3 µg/mL. After incubation for 20 minutes, the samples were washed with PBS and cell-culture medium added. Localization of fluorescence in the cells was observed with fluorescence microscopy (Eclipse TE2000).

### Multiphoton microscopy of AuNPs

For observation of AuNPs with multiphoton microscopy (modified DermaInspect; JenLab, Jena, Germany), OVCAR-3 cells were seeded in a 60 mm dish with $5 \times 10^5$ cells/mL in 4 mL medium. After cells were incubated with functionalized

AuNPs and washed with PBS, samples were diluted with medium and observed under microscopy. Different control samples were prepared, including cells incubated with nonfunctionalized AuNPs. Multiphoton signals in the wavelength range 350–600 nm were collected after excitation at 750 nm with 20 mW femtosecond-laser power.

## Delivery of dextran and cell viability

For delivery experiments, OVCAR-3 cells were seeded in a 96-well plate at a density of $2.5 \times 10^4$ cells/well in 200 mL medium. After 24-hour incubation, the medium was removed and the sample washed with PBS containing calcium and magnesium. Incubation of cells with the AuNP–antibody conjugates was carried out at 37°C with 5% $CO_2$ in the incubator. After incubation, the samples were washed again with PBS. Then, 50 mL cell-culture medium or PBS, both containing 0.4 mL 150 kDa FITC–dextran (25 mg/mL), was added to each well. In this set of samples, cells were adherent to the well bottom.

For experiments with suspended cells, OVCAR-3 cells were cultured in a flask for 48 hours. The cells were then centrifuged after trypsinization and resuspended in PBS, followed by incubation with cetuximab AuNPs for 15–20 minutes. After incubation, cells were centrifuged again and then resuspended in cell medium or PBS. The suspended cells were then transferred into a 96-well plate with ~$1.8 \times 10^5$ cells in 50 mL cell-culture medium or PBS, and 0.4 mL 150 kDa FITC–dextran (25 mg/mL) was added per well.

After irradiation with the 532 nm nanosecond laser, cells were incubated for 1 hour. Then, samples were washed with PBS and resuspended in PBS containing propidium iodide (1 µg/mL). Delivery efficiency and cell damage were evaluated using flow cytometry (FACScan; Beckman Coulter, Brea, CA, USA). Successful delivery of FITC–dextran into cells was measured by increased FITC fluorescence. Irreversibly damaged cells were stained by propidium iodide. In flow cytometry, for every sample a total of 5,000 events were examined. The results were confirmed by fluorescence microscopy in a subset of cells.

## Results

### AuNP–antibody conjugation

For AuNP–antibody conjugation, surface coverage is an important factor. To test the optimal concentration of antibody for conjugation with AuNPs, NaCl solution was added to the AuNP solutions with varying antibody concentrations. If the antibody concentration is too low, the surface of the AuNPs would not be fully covered with protein and the NPs would aggregate after NaCl addition, leading to a broadening of the absorption spectra. Ultraviolet-visible spectra of AuNP–cetuximab conjugates with different cetuximab concentrations after addition of NaCl are shown in Figure 2. Here, Ab/Au represents the antibody molecule number per AuNP. For 200 or more antibody molecules incubated per AuNP, the absorption spectrum stayed narrow after removing electrostatic repulsion by NaCl, indicating good occupancy of the surface. For the following experiments, the lowest antibody:NP ratio without broadening of the absorption peak after addition of NaCl was used (1,640 antibody molecules/AuNP).

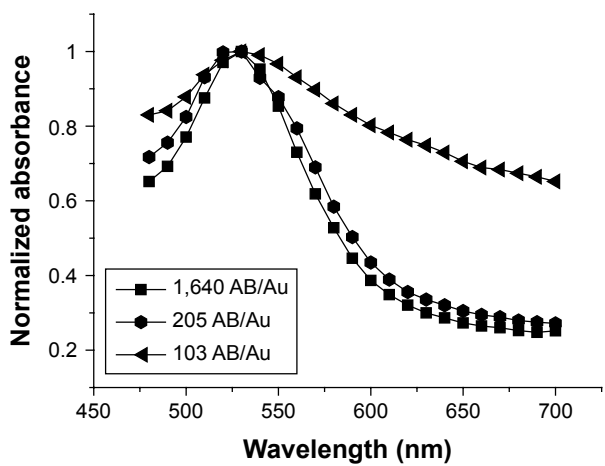

**Figure 2** Absorption spectra of cetuximab–gold nanoparticle (AuNP) complexes with different cetuximab concentrations.
**Note:** AB/Au represents the antibody molecule number per AuNP.

### Ultraviolet-visible spectroscopic measurement

After conjugation, AuNPs were also characterized with ultraviolet-visible spectrometry. Normalized absorption spectra of AuNPs and cetuximab AuNPs (1,640 Ab/Au) are shown in Figure 3. The conjugation of the antibodies to the AuNPs led to a slight red shift of the surface-plasmon resonance peak from 523 to 528 nm, due to the change in the surrounding refractive index of the AuNPs.

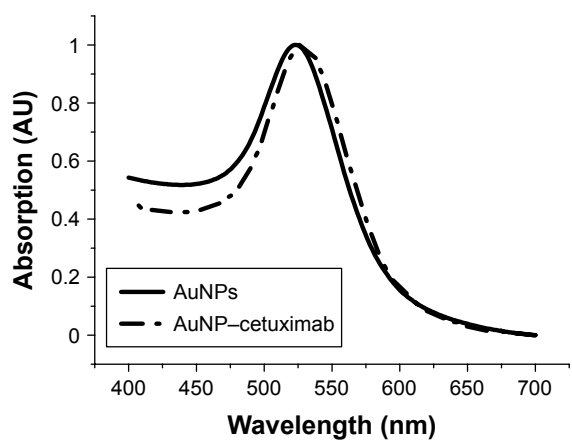

**Figure 3** Normalized absorption spectra of gold nanoparticles (AuNPs) and AuNP–cetuximab solutions.

### Microscopic imaging after silver enhancement

The binding of the cetuximab–AuNP conjugates to the cell membranes of EGFR-positive OVCAR-3 cells was tested with microscopic imaging after silver enhancement. Figure 4 shows microscopic images of OVCAR-3 cells before and after silver enhancement of AuNPs. When cells were incubated with the cetuximab–AuNP conjugates, strong silver enhancement of AuNPs on the cell membranes was observed in OVCAR-3 cells.

### Fluorescence detection of cetuximab AuNPs with a secondary antibody

For further confirmation of successful conjugation of AuNPs with the cetuximab antibody, a secondary antibody was used. While the silver reagent detects the gold particle of the conjugate, the secondary antibody is used to detect the antibody of the conjugate. Here, OVCAR-3 were incubated with cetuximab AuNPs and then with goat antihuman Alexa 546 antibodies. Without the secondary antibody, no fluorescence was observed in cetuximab AuNP-incubated OVCAR-3 cells. Figure 5 shows differential interference contrast and fluorescence images, respectively, of OVCAR-3 cells incubated with the cetuximab AuNPs and goat antihuman Alexa Fluor 546. These images show the distribution of the targeting moieties in OVCAR-3 cells and confirm the successful conjugation of AuNPs with the antibody cetuximab.

### Multiphoton microscopy and fluorescence-lifetime imaging of AuNPs

The samples were excited by 750 nm light with 20 mW for imaging. Figure 6A and B shows the multiphoton microscopy-intensity images of OVCAR-3 cells incubated with nonfunctionalized AuNPs and cetuximab AuNP. Figure 6C and D shows fluorescence-lifetime images corresponding to the intensity images of Figure 6A and B, respectively. The multiphoton signal of AuNPs targeted to EGFR was detected on the cell membranes of OVCAR-3 cells. The bright spots in the intensity images and the orange spots in the lifetime images corresponded to the distribution of AuNPs. Very few AuNPs were observed on the surface of cells after incubation with nonfunctionalized AuNPs.

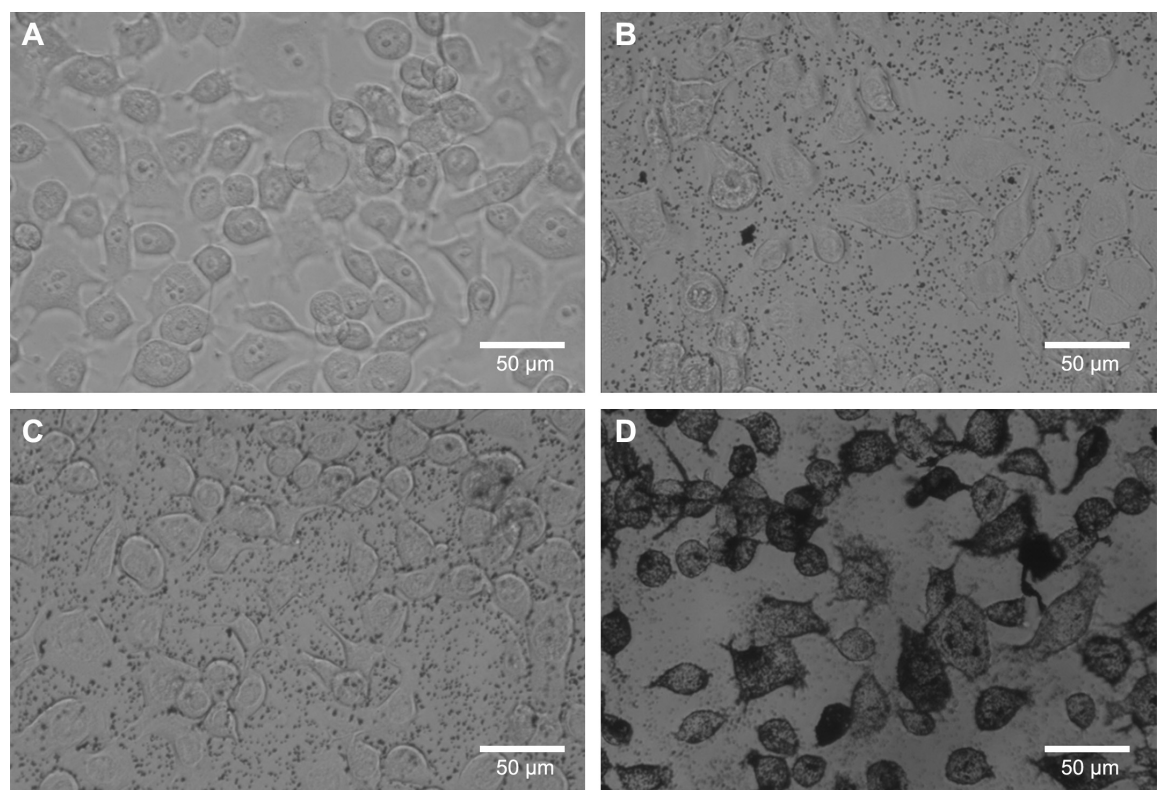

**Figure 4** Binding of cetuximab–gold nanoparticle (AuNP) complexes to OVCAR-3 cells. (**A**) OVCAR-3 cells without silver enhancement. (**B**) OVCAR-3 cells with silver reagent. Due to self-nucleation or aspecific binding, aspecific enhancement occurred at the bottom of the wells. (**C**) Cells incubated with nonfunctionalized AuNPs and silver reagent. (**D**) Cells incubated with cetuximab–AuNP conjugates and silver reagent.

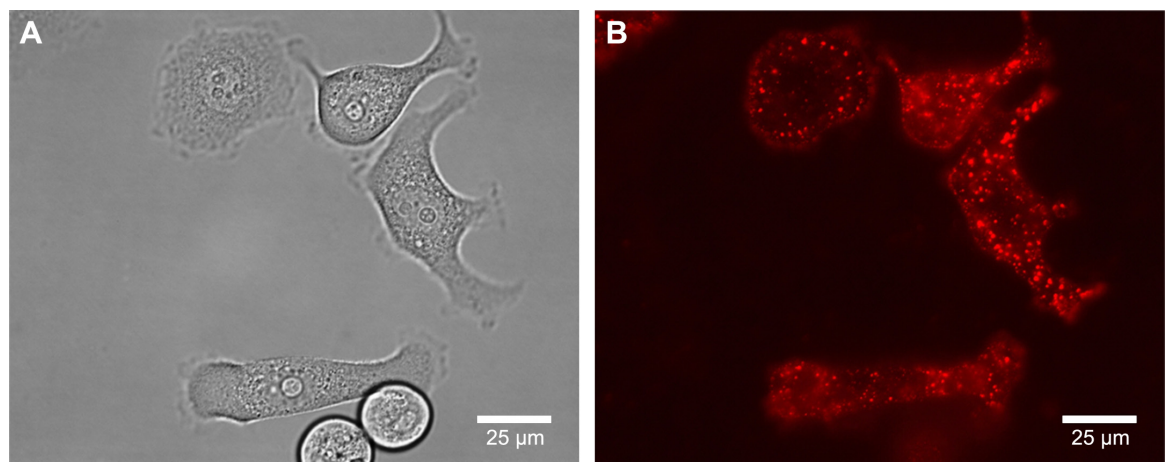

**Figure 5** Detection of cetuximab gold nanoparticles (AuNPs) with a secondary antibody.
**Notes:** Differential interference contrast (**A**) and fluorescence (**B**) microscopy of OVCAR-3 cells incubated with the cetuximab AuNPs and goat antihuman Alexa Fluor 546 secondary antibody.

## OVCAR-3 membrane permeabilization: dependence on AuNP concentration and irradiation fluence

Adherent OVCAR-3 cells were incubated with cetuximab–AuNP conjugates with concentrations between $1.4 \times 10^4$ and $9.7 \times 10^4$ conjugates/cell. Laser irradiation was carried out with fluence up to 1,900 mJ/cm². To test membrane-permeabilization efficiency, cells were incubated prior to irradiation with 150 kDa FITC–dextran, which does not cross the intact cell membrane. Figure 7A shows the variation in FITC–dextran delivery efficiency for different laser-irradiation fluence and cetuximab AuNP concentrations. Here, one can see that with irradiation below 600 mJ/cm², independent of cetuximab AuNP concentration, less than 10% of cells show effective permeability. Above 600 mJ/cm², transient permeability in 20%–30% of the cells can be observed, which strongly depends on the cetuximab AuNP concentration. Between 600 and 1,200 mJ/cm², the fraction of permeable cells increases with increasing cetuximab AuNP concentration. However, it decreases with irradiation above 1,200 mJ/cm². Also, high cetuximab AuNP concentrations resulted in less effective molecule delivery. Figure 7B shows the cell death for the same irradiation fluxes and particle concentrations as in Figure 7A. Generally, cell death percentage increased with increasing irradiation dosage and cetuximab AuNP concentration. However, the increase was higher under the condition of a higher cetuximab AuNP concentration. At high cetuximab AuNP concentrations, more cells responded to the irradiation, ie, were either permeabilized or killed compared to low cetuximab AuNP concentrations. From the results in Figure 7A and B, one can see that the highest ratio of permeabilized cell percentage over killed-cell fraction was achieved with irradiation at 1,200 mJ/cm² and a concentration of $2.17 \times 10^4$ cetuximab AuNPs/cell. Figure 7C shows the fractions of permeabilized and dead cells with different cetuximab AuNP concentrations when irradiation-pulse energy was fixed at 1,200 mJ/cm². Under optimized conditions of $2.17 \times 10^4$ particles/cell, the highest transfected:dead ratio was achieved, and about 30% cells were permeabilized. In this situation, about 20% cells were killed. With $9.7 \times 10^4$ cetuximab AuNPs/cell and an irradiation of 1,200 mJ/cm², the fraction of cells killed increased to 53%.

## Permeabilization of trypsinized OVCAR-3 cells

To investigate differences in permeabilization of adherent and suspension cells, we trypsinized OVCAR-3 cells before irradiation. After irradiation, samples were incubated for about 1 hour. Figure 8 shows the cell membrane permeability for FITC–dextran and cell death as functions of irradiation fluence with $1.8 \times 10^5$ cells per well and $4 \times 10^3$ AuNP–cetuximab conjugates per cell. Approximately 40%–65% of irradiated cells showed uptake of FITC–dextran when irradiation fluence was 400–600 mJ/cm². At the same time, the reduction in cell viability was lower than 10%. For irradiation fluence at 1,000–1,900 mJ/cm², the fraction of permeabilized cells was 70%, with only 20% cell death. Cells without cetuximab AuNPs showed neither significant membrane permeabilization nor cell death up to irradiation of 1,900 mJ/cm².

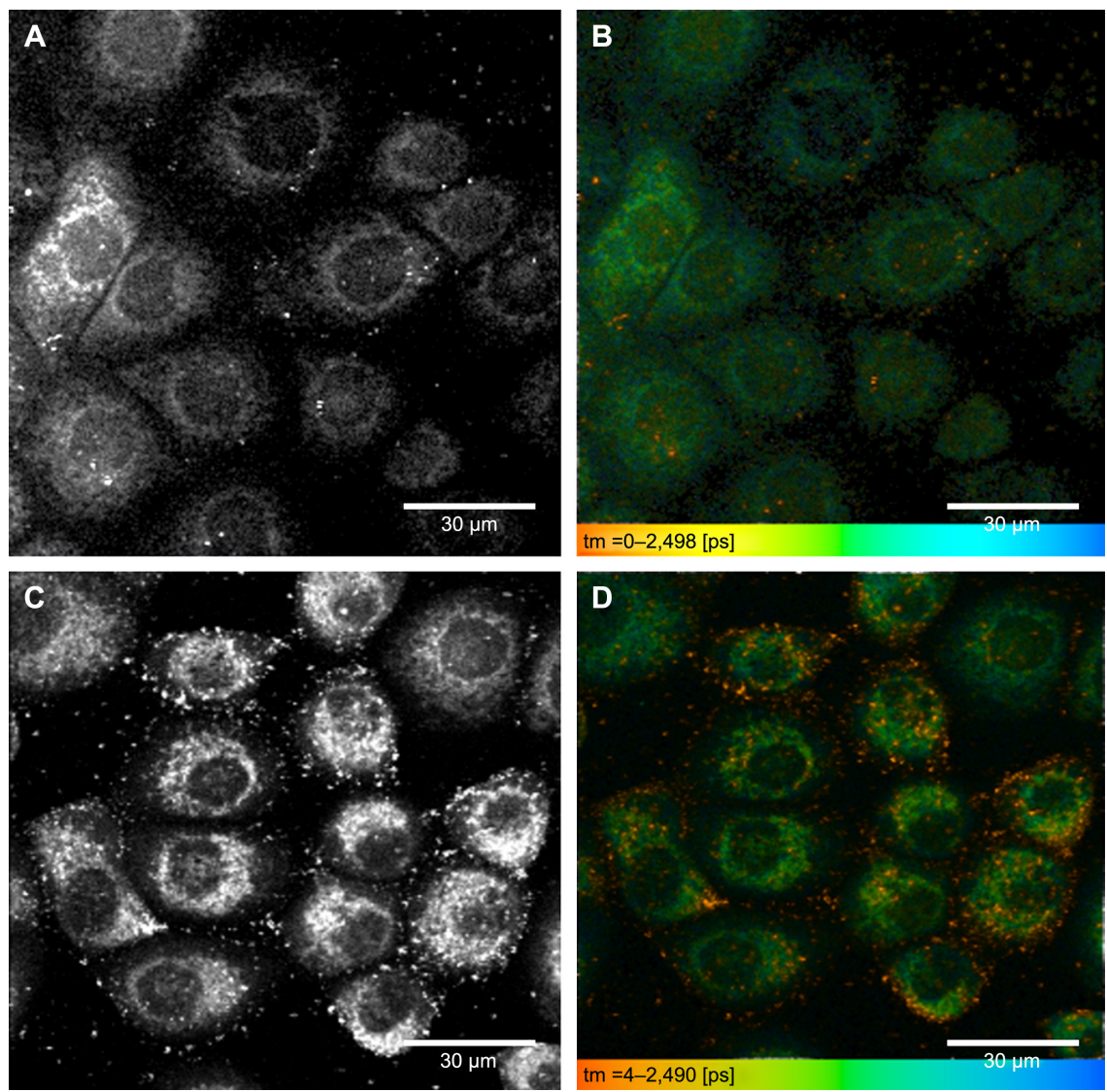

**Figure 6** Multiphoton microscopy and fluorescence-lifetime imaging of gold nanoparticles (AuNPs).
**Notes:** (**A**, **B**) Multiphoton images of OVCAR-3 cells incubated with nonfunctionalized AuNPs (**A**) and cetuximab AuNPs (**B**), respectively. (**C**, **D**) Fluorescence-lifetime images, corresponding to the intensity images in (**A** and **B**), respectively. While the autofluorescence of OVCAR-3 cells (green) showed relatively long fluorescence lifetime in the microsecond range, the AuNPs (orange) show short fluorescence lifetime in the picosecond range.

## Influence of cell environment on permeabilization

To study the influence of surrounding media on cell membrane permeability, trypsinized OVCAR-3 cells were suspended in PBS or cell-culture medium, both containing FITC–dextran. After irradiation, samples were incubated for 1 hour. Figure 9 shows permeabilized and dead cells as functions of irradiation fluence. For incubation, $4\times10^3$ cetuximab AuNPs/cell were used. When cells were suspended in PBS, 60% showed permeabilization and 10% showed cell death after irradiation with 200 mJ/cm$^2$. When irradiation fluence was increased to 1,500 mJ/cm$^2$, the fraction of transfected cells decreased to 5.48%, while the fraction of dead cells increased to 87.6%. On the other hand, when cells were suspended in culture medium, the perforation fraction reached 72.3% and the cell death fraction remained at 17.5%, even at an irradiation fluence of 1,500 mJ/cm$^2$. When irradiation fluence was higher than 1,500 mJ/cm$^2$, almost all the cells suspended in PBS died.

Figure 10 shows permeabilized and dead cell fractions when the concentration of cetuximab AuNPs was reduced to $10^3$/cell. When cells were suspended in PBS, permeabilized cells reached ~70% at 1,500 mJ/cm$^2$ irradiation fluence.

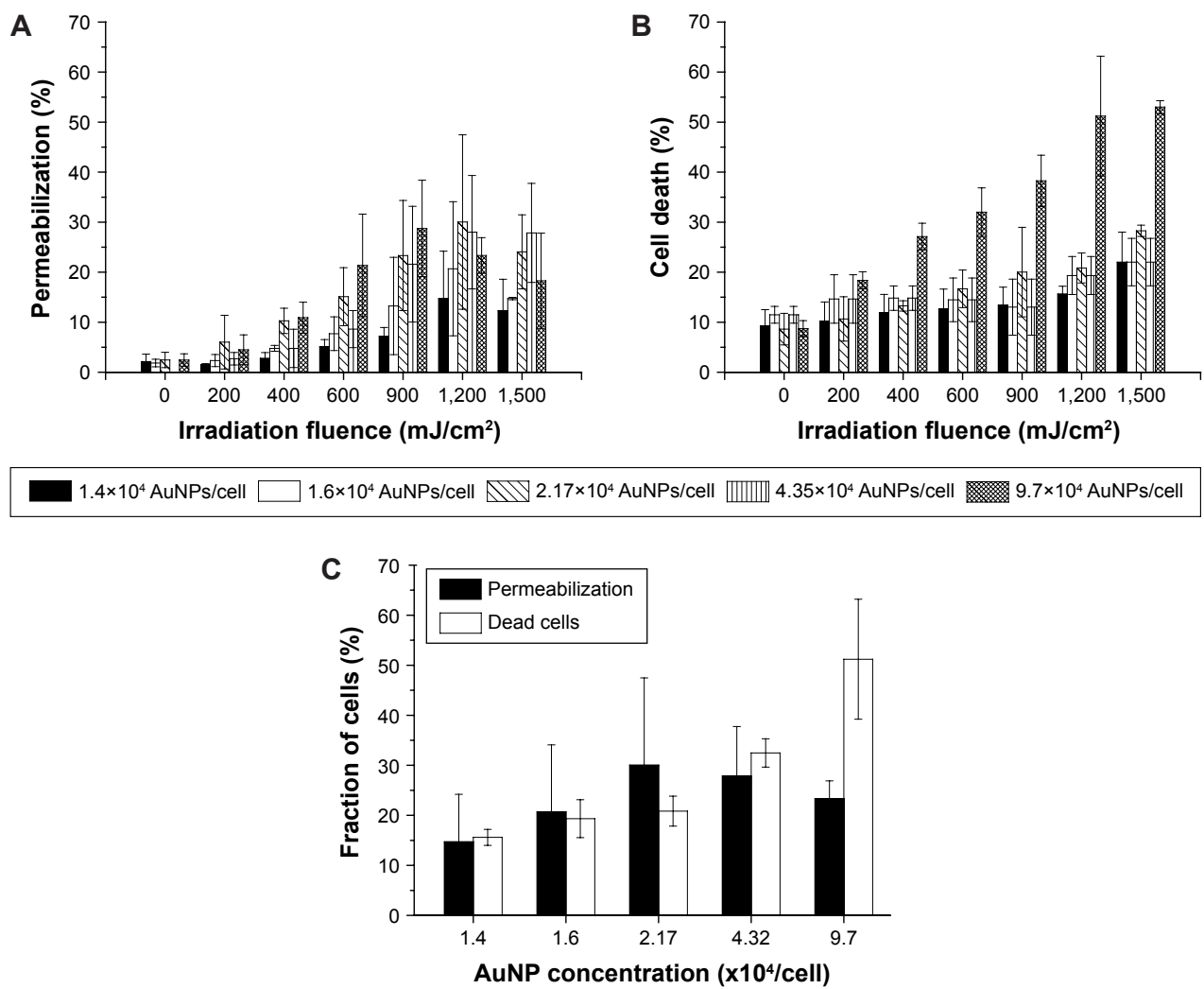

**Figure 7** OVCAR-3 membrane permeabilization depends on gold nanoparticle (AuNP) concentration and irradiation fluence.
**Notes:** (**A**) Permeabilization efficiency for 150 kDa fluorescein isothiocyanate–dextran under different laser-irradiation fluence and cetuximab AuNP concentration. (**B**) Cell death corresponding to the conditions in (**A**). (**C**) Fractions of transfected and dead cells with different cetuximab AuNP concentrations when irradiation-pulse energy was fixed at 1,200 mJ/cm$^2$.

With increased irradiation fluence, the permeabilized cell fraction decreased. When irradiation fluence was increased to 4,500 mJ/cm$^2$, fractions of permeabilized and dead cells were approximately 30% and 68%, respectively. However, when cells were suspended in culture medium, with increased irradiation fluence permeabilized cells increased to 73% at 4,000 mJ/cm$^2$, and the fraction of dead cells was similar to the control. With 4,500 mJ/cm$^2$ irradiation fluence, the fraction of permeabilized cells decreased and the fraction of dead cells increased slightly.

### Influence of AuNP-incubation time on membrane permeabilization and cell viability

Incubation time with cetuximab AuNPs was varied to analyze the influence on cell-membrane permeabilization and cell death. Cells were incubated with the conjugates for 20, 30, and 60 minutes. Figure 11 shows the cell fractions of reversible cell membrane permeability and cell death under different conditions of irradiation fluence and incubation time. For incubation of 20 minutes, the perforation fraction continuously increased with irradiation fluence up to 1,900 mJ/cm$^2$. After incubation for 60 minutes, the transient permeability fraction increased with irradiation fluence up to 600 mJ/cm$^2$ and leveled off with higher fluence. Loss of cell viability increased with irradiation fluence for all incubation times. Also, longer incubation led to a higher cell death fraction. Figure 12 shows the fluorescence-lifetime images of cell samples after different incubation times with cetuximab AuNPs. With increased incubation time, the binding of cetuximab AuNPs to cells increased.

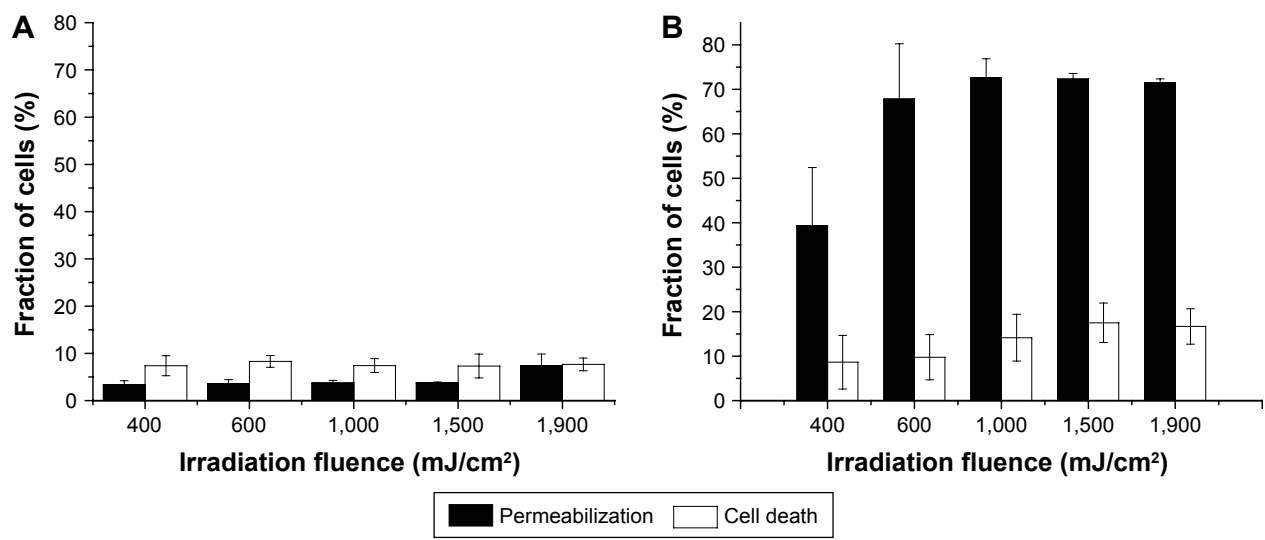

**Figure 8** Permeabilization of cells in suspension.
**Notes:** (**A**) Control; cells irradiated, but without conjugates. (**B**) Fractions of cell membrane permeability and cell death at different irradiation-fluence levels with $1.8 \times 10^5$ cells/well and $4 \times 10^3$ cetuximab–gold nanoparticle conjugates/cell.

## Discussion

Laser-based transfection is an interesting tool for effective molecule delivery into living cells. In our study, we showed transfection of adherently growing ovarian carcinoma cells (OVCAR-3) with the help of EGFR-targeted AuNPs and nanosecond pulsed-laser irradiation. Transmission electron microscopy imaging, immunofluorescence imaging, and fluorescence-lifetime imaging demonstrated the selective binding of the cetuximab AuNPs to the EGFR-positive OVCAR-3 cells. Laser irradiation led to a transient permeabilization of the membrane and to cellular uptake of 150 kDa FITC–dextran. When cells were grown adherently, we observed permeabilization in up to 30% of them, and in trypsinized and suspended cells a transfection rate of up to 70% was possible.

One reason for the difference in permeabilization rate for adherent and suspended cells may be the influence of trypsin. Trypsin is a serine protease and can cleave peptide chains. Due to the proteolytic activity of trypsin, the cellular physiology is changed,[25] and this may lead to easier permeabilization of cell membranes compared to untrypsinized cells. Also, the different distribution of EGFR molecules on the membranes of adherent and suspension cells may contribute to different permeabilization behavior.

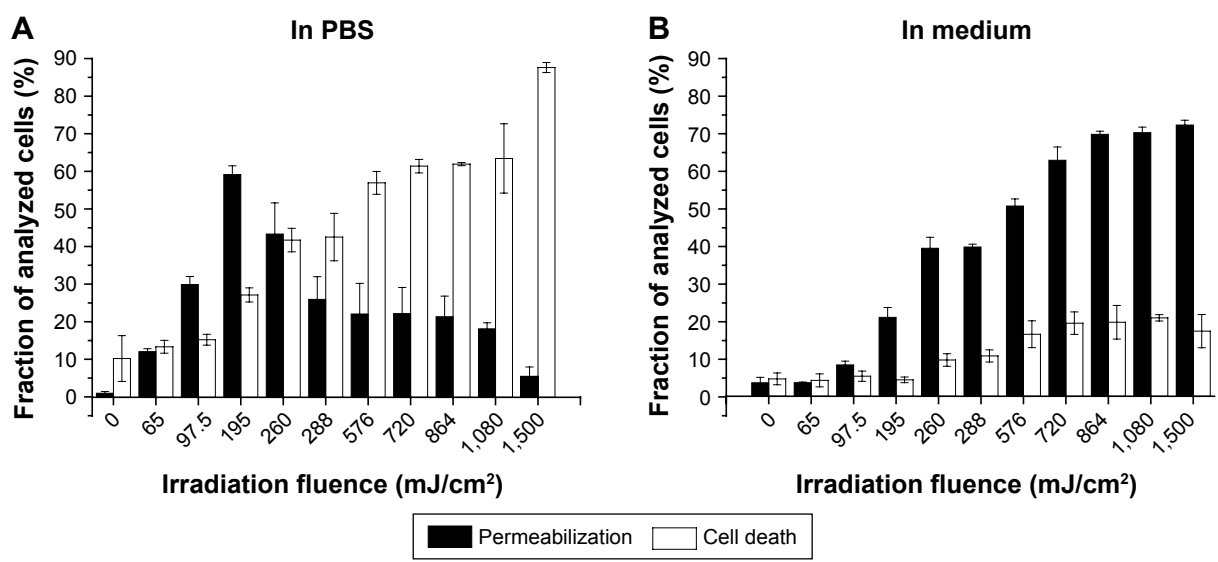

**Figure 9** Permeabilized and dead cells after incubation in PBS (**A**) and culture medium (**B**) and irradiation ($4 \times 10^3$ cetuximab–gold nanoparticle conjugates/cell used).

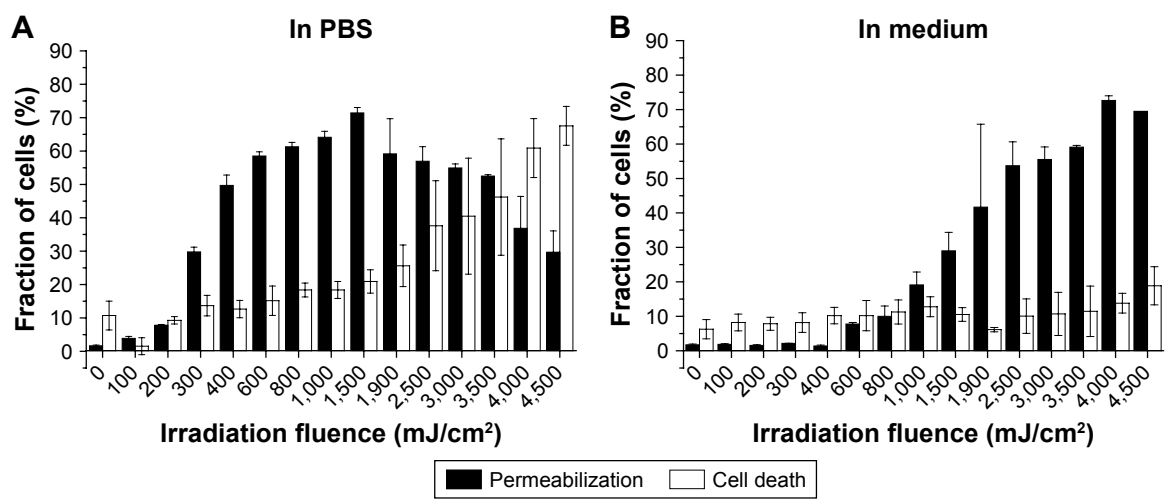

**Figure 10** Permeabilized and dead cells after incubation in PBS (**A**) and culture medium (**B**) at different irradiation fluxes (10³ cetuximab–gold nanoparticle conjugates/cell used).

Another reason may be the change in available surface area. When cells are adhered to the bottom of the well, only half of the surface area of one cell can be optoporated and receive exogenous materials. Moreover, irradiation was implemented from the well bottom, while the AuNPs bound to EGF receptors on the cell membrane facing the free medium. Therefore, laser irradiation can scatter at cell compartments, leading to a reduction in energy absorbed by the AuNPs and further to reduced transfection efficiency.

Lukianova-Hleb et al also used suspension cell lines for permeabilization. Although single laser pulses were focused on individual cells, the efficacy of GFP transfection in targeted cells reached 74%,[22] which is in a similar range to our optoporation results for trypsinized cells. For adherent cells, Lalonde et al did similar experiments and got comparable results for membrane permeabilization.[20] In their results, 26%±18% of cells were perforated with optimal fluence of 55 mJ/cm² with 532 nm nanometer pulsed-laser irradiation. However, with a femtosecond laser, transfection efficiency can reach 70%, which may be due to this laser's much higher power density. Heinemann et al and Krawinkel et al used 200 nm AuNPs with a concentration of 0.5 μg/cm² (about six particles per cell) to permeabilize membranes with picosecond and nanosecond lasers.[26,27] They demonstrated permeabilization efficiency of 88% and 85%, respectively. This is also possible, due to the size of the AuNPs, because larger size leads to higher absorption cross-section.

The key parameter for optoporation is AuNP concentration. While for suspended cells, a permeabilization rate of around 70% was achieved with an AuNP concentration of $10^3$ AuNPs/cell (Figure 10), optoporation of adherent cells required a much higher concentration of more than $1.4\times10^4$ AuNPs/cell. However, an increase in AuNP concentration is inevitably accompanied by reduced cell viability. An increase in applied irradiation additionally enhanced the effect of cell elimination.

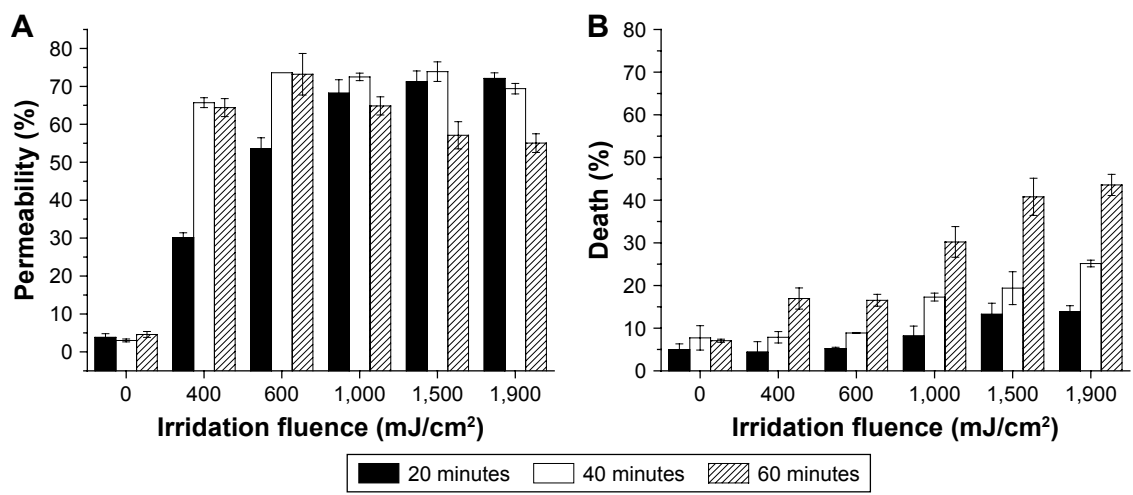

**Figure 11** Influence of cetuximab–gold nanoparticle conjugate-incubation time on cell permeabilization (**A**) and cell death (**B**).

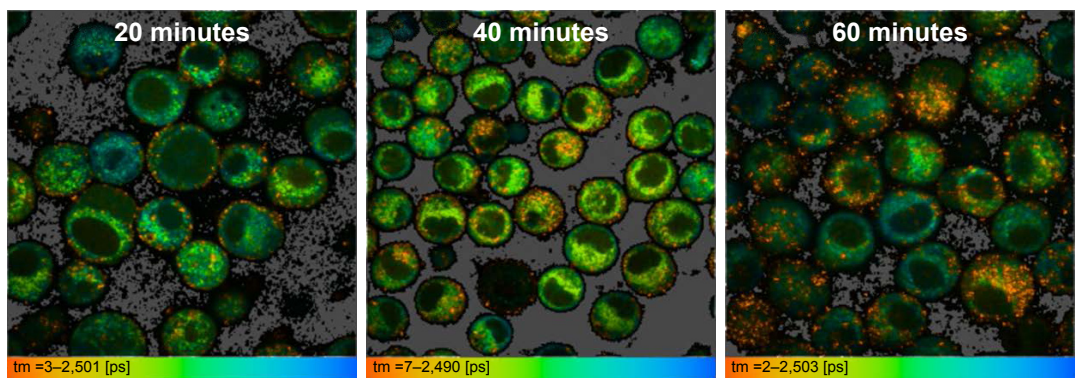

**Figure 12** Fluorescence-lifetime imaging of OVCAR-3 cells after different incubation times with cetuximab–gold nanoparticle conjugates.

Besides the parameters of AuNP concentration and irradiation, environmental changes can also affect cell permeabilization and viability. Absorption of laser energy can cause an increase in well-plate temperature, which can dissipate into the surroundings. Therefore, cells adherent to the well bottom can be affected by heating, while suspended cells will remain rather unaffected, due to cooling by the medium. However, due to the high transparency of the well bottom and the short retention time of the laser beam on the samples, heating of the well bottom and thus indirect heating of cells can be considered negligible.

Different groups have used this technique for the delivery of molecules into different cells. Their results showed that cell type was not an important factor in permeabilization efficiency. Jumelle et al delivered calcein molecules into human corneal endothelial cells with carbon NPs activated by femtosecond-laser irradiation, with efficiency of 54%.[21] Heinemann et al demonstrated delivery of two different proteins, GFP and caspase 3, into a canine pleomorphic adenoma cell line (ZMTH3) using GNOME laser transfection. Delivery efficiency of 43% was demonstrated for GFP delivery, whereas caspase 3 was delivered in more than 60% of the cells.[1] The same group also showed the delivery of Alexa Fluor 488-labeled siRNA into ZMTH3 cells with about 88% efficiency. They concluded that the GNOME approach enables cell type-independent delivery of a large variety of molecules.[27] Krawinkel et al implemented AuNP-mediated delivery of calcein molecules into primary human gingival fibroblasts (pHFIB G cells) and ZMTH using nanosecond-laser pulses. They achieved almost the same permeabilization efficiency – about 83%.[26] Baumgart et al[19] and Lalonde et al[20] used nanosecond and femtosecond lasers to introduce the fluorescent dye Lucifer yellow into human melanoma cells (MW278), with perforation rates of about 30% and 70%, respectively. They also transfected cells with DNA plasmids with femtosecond laser, and transfection efficiency (23%) was three times higher compared to conventional lipofection. Selective gene transfection of individual leukemia cells (J32) and human CD34[+] CD117[+] stem cells in vitro was performed with plasmonic nanobubbles by Lukianova-Hleb et al.[22] Transfection efficiency reached 74% and 71%, respectively.

From these different results, we can conclude that permeabilization efficiency is mostly independent of cell type. The choice of target protein on the cell membrane might be of relevance for permeabilization efficiency, since the size of the protein and the antibody-binding site on the protein determine the distance of the gold particle from the membrane, which could have an influence on membrane-permeabilization efficiency.

For laser-activated perforation, different NPs have been used, such as gold nanospheres, gold nanoshells, or carbon NPs. However, for all particles, the mechanism after pulsed-laser irradiation is similar. The laser pulse can trigger the heating of the NPs, which can induce bubble formation and influence membrane integrity mechanically.

In earlier studies, we focused on the influence of laser parameters on AuNP-mediated membrane permeabilization.[18] In this line, other groups have investigated different wavelengths[20] and different pulse durations.[1,19,21] With the results presented here, it becomes increasingly clear that not only laser parameters and particle concentrations but also the biological environment of the cell is important for the success of permeabilization. Here, we additionally varied cell medium, incubation time, and pretreatment of cells (trypsinized or untrypsinized).

We observed that for cells incubated in PBS, successful permeabilization was observed at lower irradiation fluence than cells incubated in RPMI medium. Cells in PBS also died at lower fluence than those incubated in cell-culture medium. One reason for this could be different osmotic pressure of PBS and medium, because osmotic pressure can be an additional driving force. Another reason for the different behavior in PBS or medium could be the potential of AuNPs to aggregate in PBS. Aggregated particles behave as bigger particles and would thus show similar results with lower irradiation energy. As shown in Figure 13, the cetuximab AuNPs in PBS had a wider absorption peak and higher absorption, especially in the red-wavelength range, which points to aggregation of the AuNPs in PBS.

We also compared different cetuximab AuNP concentrations per cell in PBS and culture medium ($4 \times 10^3$ and $10^3$ cetuximab AuNPs/cell). Both concentrations led to maximum permeabilization efficiency of around 60%–70% in PBS and media. However, with higher concentrations, cell permeabilization occurred at lower irradiation energies. Overall, lower particle concentrations would make permeabilization experiments more controllable. Higher concentrations could also lead to the aggregation of AuNPs to form large clusters.[28] Such large clusters of AuNP can generate bigger bubbles for more effective perforation under the same conditions of irradiation fluence.[29]

We further showed that longer incubation of AuNPs with cells led to higher association of Au conjugates in the cells, although it was difficult to distinguish from the images if the conjugates were on the membrane or inside the cells. Aspecific endocytotic uptake has been demonstrated for AuNPs,[30] and also receptor-mediated uptake via EGFR binding has been shown.[31] Increased incubation time led to permeabilization at slightly lower energies and to a more pronounced shift of cell death to lower irradiation energy. However, the cell death fraction increased significantly with incubation time, as shown in Figure 11B. Therefore, when incubation is longer than 60 minutes, the number of dead cells may further increase and dominate the photoporation process. This can be explained by an increased number of conjugates bound to the cell membrane and also an increased number of conjugates that are taken up into the cells via endocytosis. As our results show, a higher number of conjugates led to increased cell death at the same irradiation fluence. Moreover, internalized AuNPs are less effective in cell inactivation compared with AuNPs remaining on the cell membrane.[32] We can speculate that AuNPs inside cells are less efficient in permeabilization of the membrane, but destroy internal cellular structures for localized thermal damage.[33] The evolving cavitation bubbles around the particles can have diameters up to the micrometer range, and could damage intracellular structures and compartments, like endo/lysosomes, mitochondria, or the endoplasmic reticulum. For application of this technique for therapeutic molecule delivery, incubation time is a critical factor for optimal permeabilization results.

This laser-assisted molecule delivery in combination with cell elimination at higher irradiation energy also represents an interesting tool for selective manipulation of individual cells. This has been investigated for use in stem-cell differentiation and tissue engineering.[34] Cells that are differentiating in an unpreferred direction can be eliminated with higher laser energy. This elimination is very selective at the cellular level,[35] while differentiation could be driven additionally by optically controlled permeabilization for uptake of cytokines, for example. Individual cell types can be addressed by the use of different antibodies on the AuNP surface addressing different membrane proteins.

## Conclusion

We have demonstrated membrane permeabilization of adherent and suspended OVCAR-3 cells after incubation with cetuximab AuNPs and irradiation with a nanosecond laser at 532 nm wavelength. In particular, we showed the effects of various factors, including AuNP conjugation (antibody linkage or not), AuNP concentration, irradiation fluence, cell condition (adherent or suspension), cell environment (in PBS or culture medium), and cell-incubation time with AuNP conjugates on cell permeabilization and death. The optimal perforation efficiencies for adherent and trypsinized cells were about 30% and 70%, respectively. Perforation efficiency was sensitive to the incubation time of cells with conjugates. The perforation rate was also strongly affected by the environment of cells during irradiation. The highest perforation efficiency was achieved with lower irradiation fluence or lower AuNP concentration when cells were in PBS. The cause of this difference deserves further investigation. The optoporation results show great promise for macromolecule delivery and gene transfection and may be useful for optimization of AuNP-mediated laser molecule delivery.

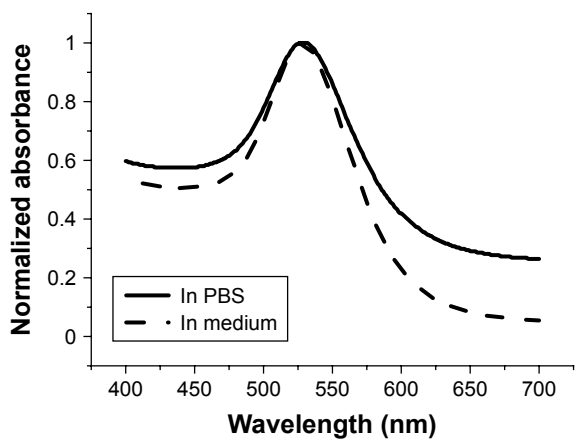

**Figure 13** Normalized absorption spectra of cetuximab–gold nanoparticle conjugates in PBS and medium.

## Acknowledgments

We acknowledge Barbara Flucke and Astrid Rodewald for their assistance during this study, and we thank Professor CC Yang of Taiwan University a lot for his help. This work was supported by the German Ministry of Education and Research (grant 13N11832), the German Research Foundation (grant RA1771/3-1), the European Union project ENCOMOLE-2i (Horizon 2020, ERC CoG 646669), and the National Natural Science Foundation of China under grants 61575156 and 61335012.

## Disclosure

The authors report no conflicts of interest in this work.